\documentclass[twocolumn,showpacs,showkeys,preprintnumbers,amsmath,amssymb]{revtex4}
\usepackage{graphicx}% Include figure files
\usepackage{dcolumn}% Align table columns on decimal point
\usepackage{bm}% bold math

\begin{document}

%\preprint{APS/123-QED}

\title{Molecular dynamics simulations of intensive plastic deformation}
% Force line breaks with \\

\author{L.S.Metlov}
\email{metlov@mail.donbass.com, metlov@ukr.net}
\affiliation{Donetsk Physico-Technical Institute, Ukrainian
Academy of Sciences,
\\83114, R.Luxemburg str. 72, Donetsk, Ukraine}

\thanks{This work was supported by the Fond of Budget Researches of National
Academy of Ukraine (topic 108) and by Donetsk Innovation Center
(provision of computer and internet)}

\thanks{Author express great thanks to Yu. V. Eremeichenkova
for help in translation.}

\date{\today}% It is always \today, today,

\begin{abstract}
Kinetics of dislocations is studied by means of computer
simulation during intensive plastic deformation. The dynamical
effect in the form of soliton-like wave of sharply disrupted
interparticle bonds is observed. Along with it, micropores similar
to steam bubbles at water boiling are formed. After some
deformation the solid takes an ideal structure again owing to
"emission of the bubbles". The dislocations in this state (with
micropore) can form a grain boundary in the case of
nano-structural materials. The nanoscopic object of rotational
nature is observed at uniaxial intensive deformation. Rotation of
some volume of sample which realized with phase transition
scenario was observed by intensive shear deformation.
\end{abstract}


%Use showkeys class option if keyword
\maketitle

\section{Introduction}

A new methods of intensive plastic deformation for achievement of
nano-structure in metals had been developed \cite{PMVM02}. It is
common knowledge that the main vehicle of plastic deformation is
dislocation. In recent papers new vehicle of plastic deformation
had been investigated widely for metal glasses. It is so called
shear transformation zone (STZ) \cite{EdOa89}, \cite{FL98},
\cite{Falk99}, \cite{Lem02}. On other hand, there is a large body
of research of different dynamical objects in non-linear chains.
They are moving breathers in different non-linear systems
\cite{CrDauRT} and kinks in acoustic chains \cite{Metlov},
\cite{MeEr02}. The main property of these objects is spontaneous
concentration energy in the large excitations at cost of suppress
of weak ones. It is possible to assume that such concentration in
three-dimensional case can lead to rearrangement of internal
structure of solid during intensive deformation.

\section{Constitutive relations}

For the check of mentioned above assumption numerical experiment
on two-dimensional model was performed by means of MD-simulation.
Model system of 18*18 particles with Lennard-Jones potential is
used:

  \begin{equation}\label{a1}
    U_{ijlk}=\frac{A}{6}(\frac{B}{2r_{ijlk}^{12}}-\frac{1}{r_{ijlk}^{6}}),
  \end{equation}
 where $r_{ijlk}=\sqrt{(X_{ij}-X_{lk})^{2}-(Y_{ij}-Y_{lk})^{2}}$ - is the
 distance between the particles numbered by $i,j$ and $l,k$
 with Cartesian coordinates $X_{ij}, Y_{ij}$ and $X_{lk}, Y_{lk}$.
 Indexes $i, l$ numerate atoms in the lattice along the $Y$-direction,
 $j, k$ do the same along the $X$-direction. Let use a unit in
 which the constants are equal to $A=0.0025 J_{c}m_{c}^{6}$ , $B=1 m_{c}^{6}$,
 where $J_{c}$ and $m_{c}$ are conventional units of energy and length
 respectively.  We put the mass of a particle to be $M=0.01 kg_{c}$,
 where $kg_{c}$ is conventional  unit of mass. The binding energy of pair
 interaction is found to be $E_{b}=A/12B=0.00020833 J_{c}$ and equilibrium distance
 is $r_{0}=\sqrt[6]{B}=1 m_{c}$. In case of real material it is necessary to
 perform the re-computation with account of real values of $E_{b}, r_{0}$ and the
 mass $M$. For example, in the case of copper $E_{b}=0.5493\ast10^{-19} J$, $r_{0}=2.66
 \ast10^{-10} m$, $M=1.0541\ast10^{-25} kg$.   Then, $1m_{c}=2.66*10^{-10} m$,
 $1 kg_{c}=1.054*10^{-23} kg$, $1 s_{c}=0.532*10^{-13} s$. The
 time unit have the same order as the period of small vibrations $T=2\pi{m}/U_{ijlm}''$.
 Time step for simulation is chosen to be $\triangle{t}=0.18 s_{c}$.

\section{Uniaxial loading}

The first series of real time computer simulations is fulfilled
for uniaxial loading of a crystallite sample. The system is placed
on rigid platform constructed from the atoms identical with those
in the sample (the lowermost line of atoms on the figures). The
rigid atomic line (plane in a three-dimensional case) constructed
from the same atoms moves downward with a constant velocity
$2.778*10^{-4} m_{c}/s_{c}$ (the uppermost line of atoms on fig.
\ref{f1}). Initial atom configuration is hexagonal lattice with
interatomic distance $r_{0}=1m_{c}$.

In the first stage the usual uniaxial elastic deformation goes
(Fig.\ref{f1}a).
\begin{figure*}
  %\includegraphics[width=5.5in]{Probe}\\
  \includegraphics [width=7 in] {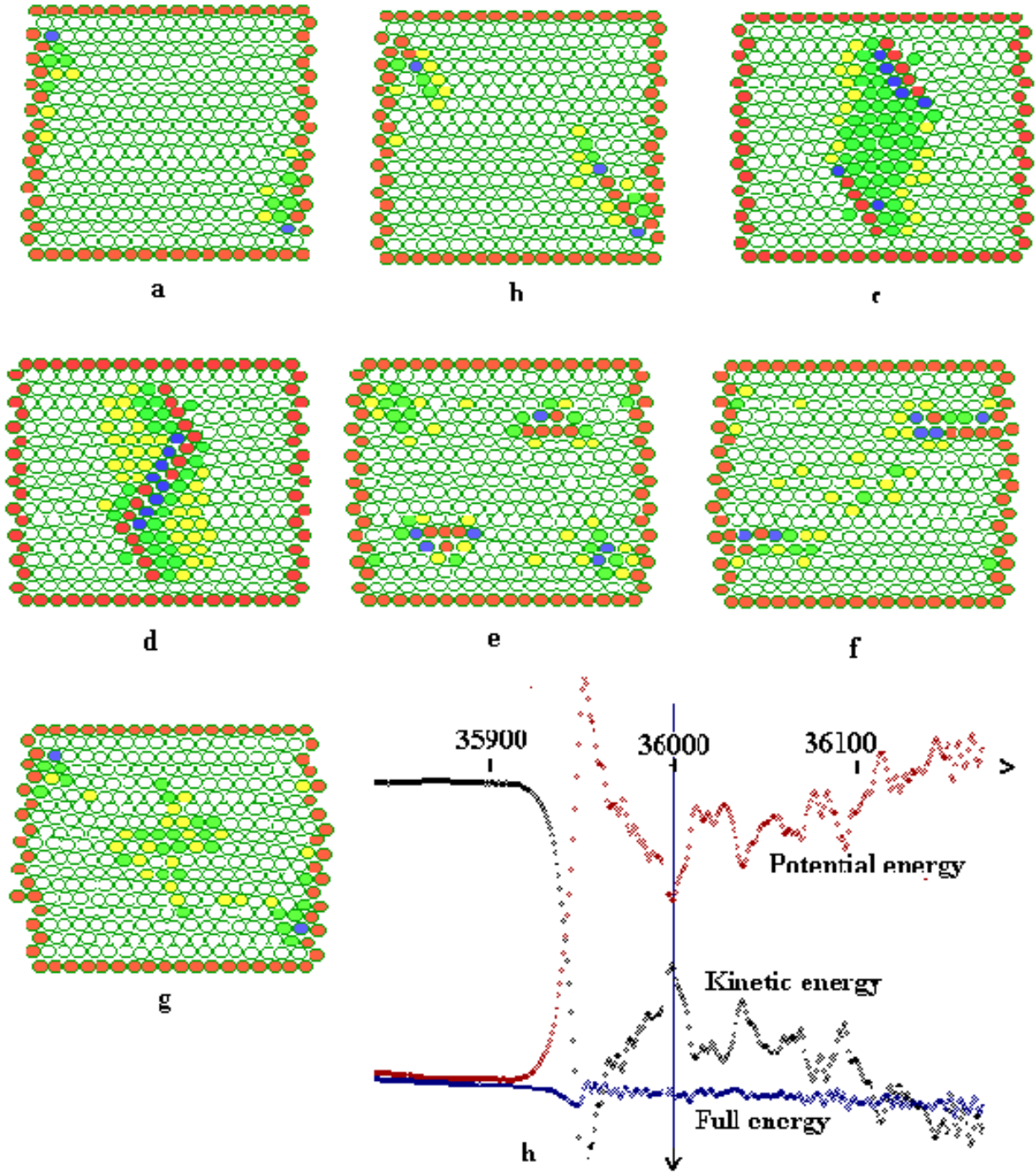}
  \caption{\label{f1} Evolution of crystallite structure at uniaxial loading:
  a - ideal crystal structure at time step 27007,
  b - birth of dislocation pair at 29253,
  c - full formed bidislocation pair at 35 35019,
  d - lightning-like breakthrough between bidislocations at 35923,
  e - turned bidislocation pair at 36042,
  f - pressing out of the bidislocations from the sample at 36117,
  h - graphics of total potential energy of the system and kinetic and full ones during
  the turn of the bidislocations. Average potential energy per atom is equal to
  $-0.00132J_{c}$.
  Yellow circles present the atoms with potential energy $0.00002J_{c}<U<0.00004J_{c}$,
  green circles - with $0.00004J_{c}<U<0.0001J_{c}$,
  blue circles - with $0.0001J_{c}<U<0.00015J_{c}$,
  red circles - $0.00015J_{c}<U<0.00038J_{c}$,.
  Potential energy of the particles on free lateral surfaces is equal $0.0006J_{c}$.}\label{f1}
\end{figure*}
Further, two microscopic defects occur almost simultaneously in
the left upper corner of the lattice and in the right down one
(Fig.\ref{f1}b). Their Burgers vectors have identical direction
(at a $120^{\circ}$ angle with horizontal axis) and opposite
signs. These defects are similar to usual dislocations
\cite{BaSeMy00}. However each of them has two and only two
inserted atomic lines. They can't have one inserted atomic line
since atomic configuration doesn't permit it owing to very large
energy of the defects. Two inserted lines and glide line are
oriented at $60^{\circ}$ angle each to another. Both of the
inserted lines always lie on one and only one side of the glide
line. It is pertinent to call a defect with two inserted atomic
lines as bidislocation. Along with two-dimensional lattice, such
defects may be observed in three-dimensional case on the planes of
closely packed layers (planes $\{1,1,1\}$). In general case of
hexagonal lattice six type of bidislocation having Burgers vectors
with the angles $0, \pi, \pm\pi/3, \pm2\pi/3$ may take place. In
complex plane these vectors are expressed as

\begin{equation}\label{a2}
    \pm{a_{0}}, a_{0}\exp(\pm\frac{\pi}{3}i),
    a_{0}\exp(\pm\frac{2\pi}{3}i),
\end{equation}
where $a_{0}$ - is equilibrium distance in the crystallite.

In these terms, arisen bidislocations have following Burgers
vectors:

\begin{equation}\label{a3}
    a_{0}exp(-\frac{\pi}{3}i)\quad and \quad a_{0}exp(\frac{2\pi}{3}i).
\end{equation}

Under following loading the bidislocations move along their glide
lines for as long as they reach the rigid lines of atoms (Fig.
\ref{f1}c). Then, the bidislocations stop for some time because
the rigid lines play the role of obstacles for further moving.
Owing to this retardation, the bidislocationons form the pair with
zeroth total Burgers vector. In this case the rotation of the
region  between the bidislocations takes place. This region has
approximately circle form. The bidislocations lie exactly on
diametrically opposite sides of the circle boundary. In this
nanoscopic region the orientation of atomic lines (planes) along
Burgers vector changes continuously. Owing to this property,
observed pair of bidislocations may be simplest two-dimensional
model of displanation \cite{Konst}, \cite{KonPrDo}. The
displanations are known as defects of mesoscopic level. It is
clear from presented experiment that simplest modification of the
displanations has nanoscopic nature.

Stopping of the bidislocations creates difficulties for repairing
of ideal crystal symmetry. As a consequence, local concentration
of elastic energy around defects takes places. It is clear that
for finishing of symmetry repairing gliding lines must turn to the
direction along rigid lines of atoms. What scenario does the
system choice for it?

At the time of 600 time steps flashover-like puncture of the
region between the bidislocations occurs (Fig. \ref{f1}d). After
it both the bidislocations are turned (Fig. \ref{f1}e) and pressed
out from the lattice in the direction parallel to rigid lines
(Fig. \ref{f1}f). Thereafter, the sample takes ideal crystal
structure. Then, with continuing of loading the processes are
qualitatively repeated.

It is known that different transformations of dislocations can be
regarded as chemical \cite{SKZ01} or, better still, as elementary
particle reactions. From this standpoint, in the case of
two-dimensional hexagonal lattice any bidislocation may be
considered as an elementary particle with vector charge (2) or as
three particles (like quarks) with different own polarization and
with internal coordinate (as spin). Let us denote the set of
"elementary particles" as
\begin{widetext}
\begin{equation}\label{a4}
    \pi_{1}^{\pm}=\pm{a_{0}},
    \quad \pi_{2}^{\pm}=\langle{a_{0}}\exp(\frac{2\pi}{3}i),
    a_{0}\exp(-\frac{\pi}{3}i)\rangle ,
    \quad \pi_{3}^{\pm}=\langle{a_{0}}\exp(\frac{\pi}{3}i),
    a_{0}\exp(-\frac{2\pi}{3}i)\rangle.
\end{equation}
\end{widetext}

Then  the transition with rotation of glide lines (fig. 1.d-f) can
be written in the form of reaction:

\begin{equation}\label{a5}
    \pi_{2}^{+}+ \pi_{2}^{-}\rightarrow\pi_{1}^{+}+ \pi_{1}^{-}.
\end{equation}

Therewith, the law of conservation of charge is fulfilled:

\begin{equation}\label{a6}
{a_{0}}\exp(\frac{2\pi}{3}i)+a_{0}\exp(-\frac{\pi}{3}i)=a_{0}-a_{0}=const=0.
\end{equation}

The law of conservation of charge momentum is fulfilled since the
distances between glide lines pre and post-reaction are equal (7
lines between them on the fig. \ref{f1}d,f). The law of
conservation of energy is fulfilled since the system is Hamilton
one. As own energy of each of two bidislocations pre and
post-reaction doesn't change (in consequence of their identic
structure) accumulated elastic energy almost completely turns into
kinetic energy (relaxation of stress).

Regarding all the pictures presented in the fig. \ref{f1}, one can
note that bidislocation nucleus isn't localized in the point, but
it is continuously smeared along the glide line in the region of
several atomic sizes. The discrepancy of crystalline atomic lines
begins with the distortion of them at bidislocation edges and
increases gradually toward  center. Near a glide line the chain of
atoms viewed from the side of more rarefied region of crystallite
is stretched. Let call this atomic chain as "stretched" chain. The
chain viewed from the opposite side of the glide line is termed as
"compressed" chain. The atoms in the stretched chain are localized
in energetically unfavorable positions to ones in the compressed
chain. However, the atoms are fixed in these positions owing to
effective interaction with those in their own chains and with
nearest atoms of the rest lattice.

Let set as 1 equilibrium interparticle distance in the region far
from the bidislocations. Then, in the compressed chain
interparticle distance is equal to 1.1 and to 0.95 in stretched
chain. Moreover, all the simulations exhibit the next trend. In
the chains parallel to compressed and stretched ones and viewed
from the compressed chain interparticle distance takes its
asymptotic value 1 rapidly, already in the first chain.
Alternately, in parallel chains viewed from the stretched one
several chains are markedly stretched. Thus, the region of
compression is more localized then the region of rarefying. Owing
to this, potential energy of any particle in the compressed chain
is more large than one in asymptotic zone. This peculiarity is the
main sign for structure defects of any nature and permits to
separate bidislocations immediately during computer simulation.
Moving colored bidislocations against the background of the rest
atoms is very nice picture. Average potential energy per atom in
the compressed chain consists of 60-70 percent of one on free
surface of the sample. Potential energy of whole the system in
initial state of lattice is chosen as zeroth level.

\section{"Knife" loading}

More interesting result is obtained in the second series of
computer simulation for cutting the sample by three-atom knife
(fig. \ref{f2}).
\begin{figure*}
  %\includegraphics[width=5.5in]{Probe}\\
  \includegraphics [width=6.5 in] {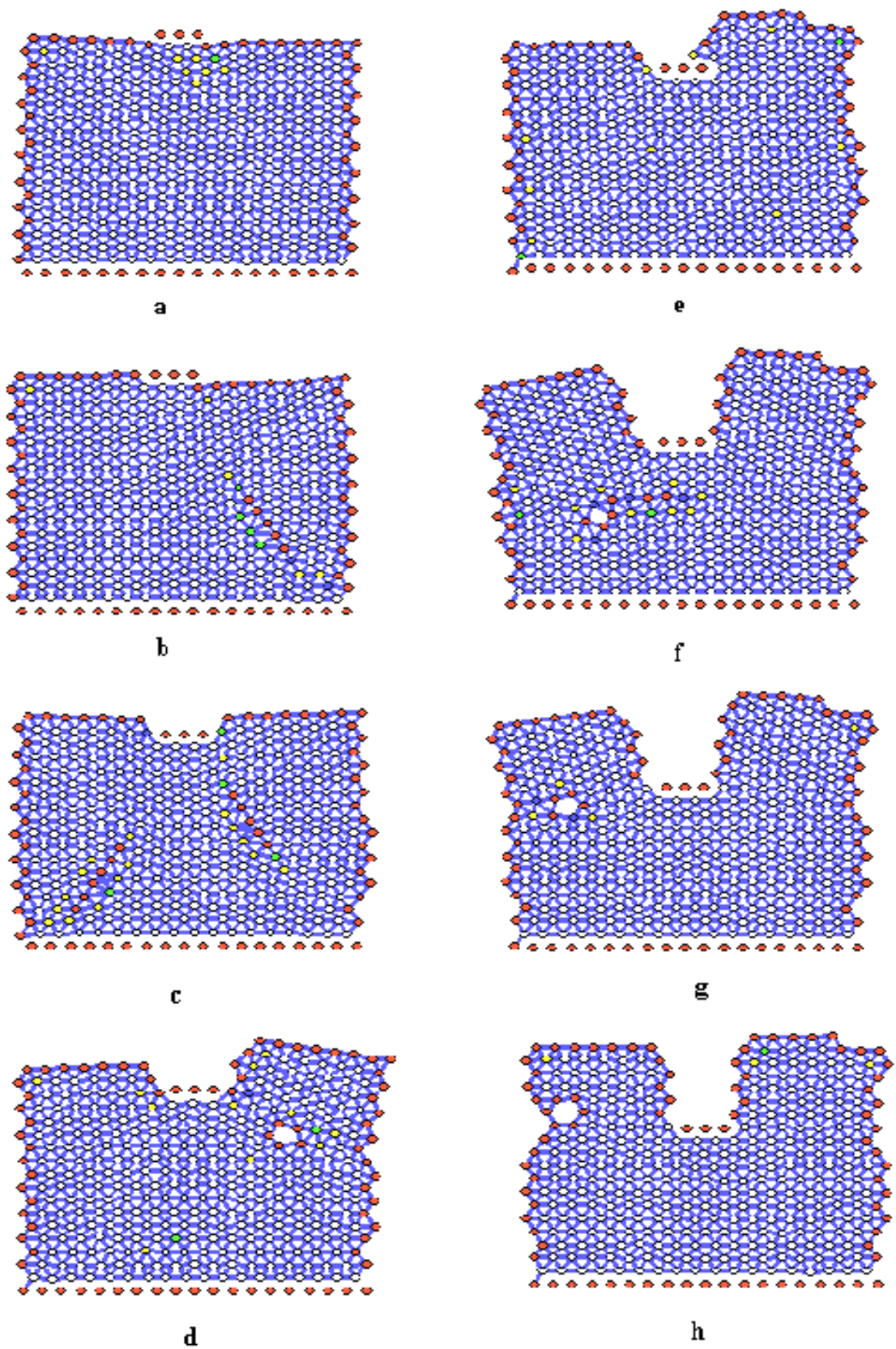}
  \caption{\label{f2}
 Evolution of crystallite structure at cutting by three atom knife:
 a - ideal crystal structure in beginning stage,
  b,c - formation of usual bidislocation,
  d, f-h - formation of bidislocation with micropore,
  e - renewal ideal crystal structure after first "bubble".}
\label{f2}
\end{figure*}
Three atoms only are left from moving upper rigid line of atoms.
Other parameters of the experiment remain as in the Sec. III.

In the first stage, in general, the picture qualitatively looks
like to one obtained in the previous numerical experiments.
Elastic stage of deformation isn't so interest. Further, the pair
of bidislocations is created in the upper left and right corners
of the crystallite. Their glide lines are turned by $60^{o}$ each
to another. The bidislocations occur in the region under
three-atom indentor and move downward. Then, the bidislocations
turn and they are pressed out from the sample.

Interesting phenomenon is observed at such pressing out (Fig.
\ref{f2}). The defect displayed on the figure has two additional
lines and Burgers vector as the bidislocation does, but its form
sufficiently differs. The stretched chain has gap in the center of
bidislocation nucleus. That is, binding among centered atoms is
broken and micropore arises.

Owing to the micropore, the bidislocation becomes more local and
more sharp. Atomic lines in the regions separated by the normal to
glide line are inclined at a big angle each to other. One can
regard the line which separates such regions as an element of a
boundary between grains.

During numerical experiment new bidislocations  with micropores
arise and disappeare again and again. Whole the picture is similar
to formation of steam bubble at water boiling. Owing to this
"emission of the bubbles", the solid  take on ideal structure
again after some deformation.

On the other hand, the movement of a such object through the
lattice is similar to kink-like excitation in acoustic chain.
Wave-like disruption and repairing of bonds between atoms give
rapid pressing out of defects from a sample.

Thus, at intensive influence on metal it may release not only by
well known shear deformation, but by wave-like (or
excitation-like) bond disruption, which moves rapidly enough. One
can assume that the micropore, namely, is the sign of new state of
bidiislocation, which arises just during intensive deformation. In
three-dimensional case the micropores can be spherical or elliptic
bubbles. After stopping of deformation they disappear through the
"condensation" into usual dislocations. Probably, it is difficult
to detect the bidislocations with micropores in experiment as
separate objects. Usual X-ray diffraction analysis isn't suitable
for investigation of such states and structure re-arrangements for
two reasons. By the first, such analysis is fulfilled much later
after finish of intensive plastic deformation. By the second, the
solid in such process can be in non-equilibrium amorphous state
with large widening of X-ray spectrum line. For detection of such
states and its changes, it is necessary to execute X-ray analysis
just during intensive plastic deformation or, maybe, immediately
after it.

\section{Shear loading}

A great amount of such objects occurs in the case of intensive
deformation by shear scheme. This scheme is realized in the
following way: the sample consisted from 28*28 particles is placed
in rigid two-dimensional box constructed from atoms identical with
those of sample. The box is deformed by horizontal movement of
upper and down rigid lines with constant velocity.

The created dislocations with micropores form boundary separated
regions with different orientation of crystalline atomic lines
(Fig.\ref{f3}). Such boundary can play important role for forming
of nano-structural state by means of the methods of large shear
deformation.

\begin{figure*}
  %\includegraphics[width=5.5in]{Probe}\\
  \includegraphics [width=6.5 in] {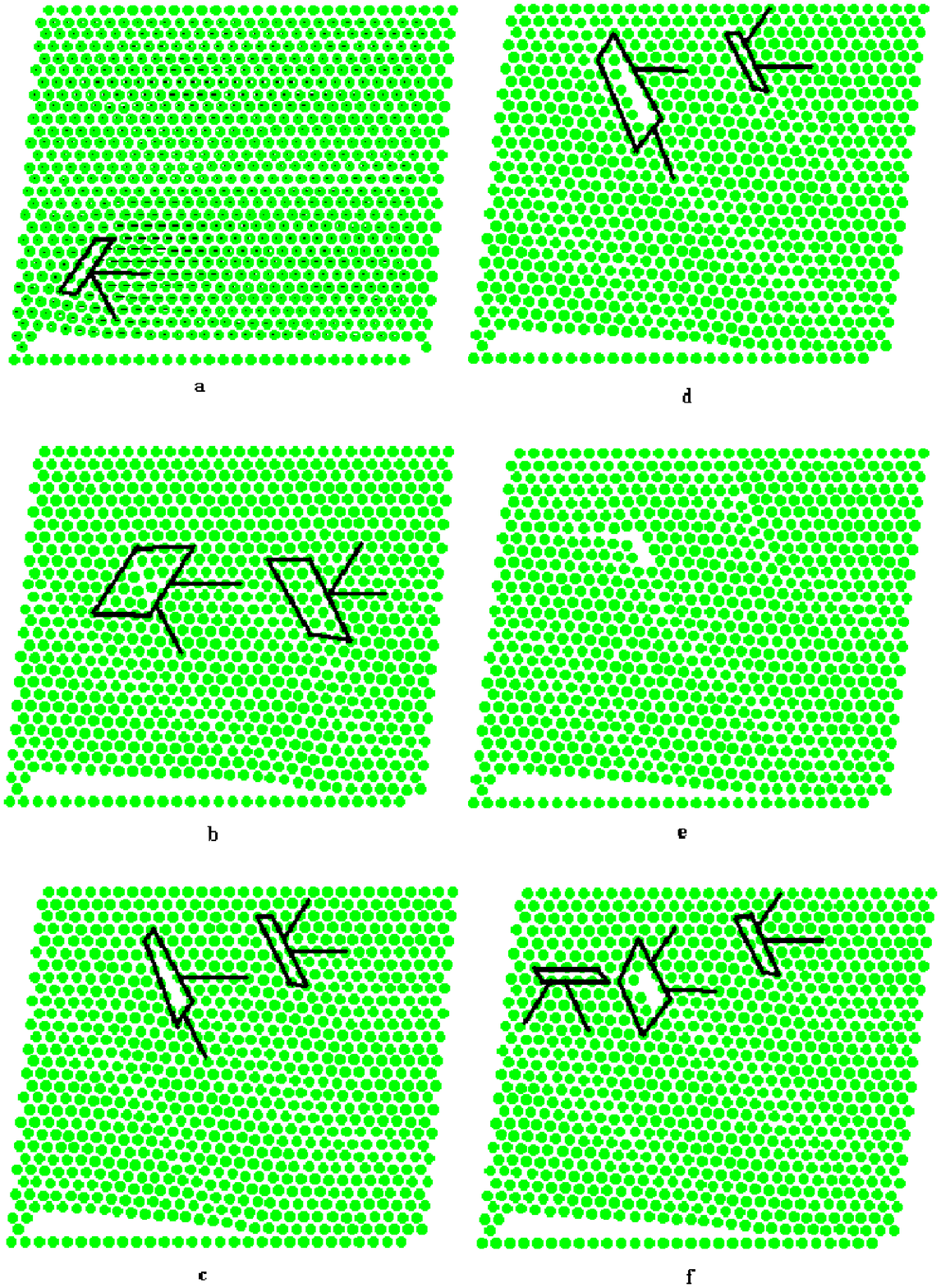}
  \caption{\label{f3}
Evolution of crystallite structure with shear deformation:
  a - birth of first bidislocation, b-d - moving of two defects,
  e - bidislocation decay with birth of two new bidislocation,
  f - system after bidislocation decay.}
\label{f3}
\end{figure*}

At the beginning stage of deformation single bidislocation is
borne in the left down corner of the sample (Fig. \ref{f3}a). It
has the charge  $a_{0}\exp(-2\pi{i}/3)$ and moves along glide line
toward the upper boundary. At the following stage, the second
bidislocation with the charge $a_{0}\exp(-\pi{i}/3)$ comes into
being and moves along glide line toward the upper boundary too
(Fig. \ref{f3}b). The glide lines of both bidislocations are
crossed at the same point exactly on the upper boundary. With the
lapse of time the first bidislocation takes the form of microcrack
(Fig. \ref{f3}c) and preserves this form for a time (Fig.
\ref{f3}d). In the following lapse the first bidislocation begin
to take the form of sharp bend (Fig. \ref{f3}e). Upon it, pair of
bidislocations arises instead the first one. One of them has the
charge $-a_{0}$ and another has the charge $a_{0}\exp(-\pi{i}/3)$.
Their total charge is equal to $a_{0}\exp(-2\pi{i}/3)$, i.e. it is
equal to the charge of the initial first bidislocation. Such
transformation can be written as decay reaction in the form:

\begin{equation}\label{a7}
    \pi_{2}^{-}+ energy\rightarrow\pi_{1}^{-}+ \pi_{3}^{-}.
\end{equation}
where the second term in the left part of equation is the energy
of stress field which causes the birth of additional "particle".
The rest of stress field energy transforms into thermal energy. As
the result, almost full relaxation of elastic stress take place.

The dislocations with micropores form boundary separated regions
with sharply different orientation of crystalline atomic lines
(fig. \ref{f4}). Such boundary can play important role for forming
of nano-structural state by the methods of large shear
deformation. Single, but very interesting result had been fixed in
computer experiment when rotation of certain volume by scenario of
phase transition takes place. At some deformation the boundary is
formed  as a chain of bidislocations with micropores. At following
loading it moves steadily until escape from the sample. In this
case the boundary separates phases of identical nature, but with
different orientation of crystalline atomic lines only.

\begin{figure*}
  %\includegraphics[width=5.5in]{Probe}\\
  \includegraphics [width=7in] {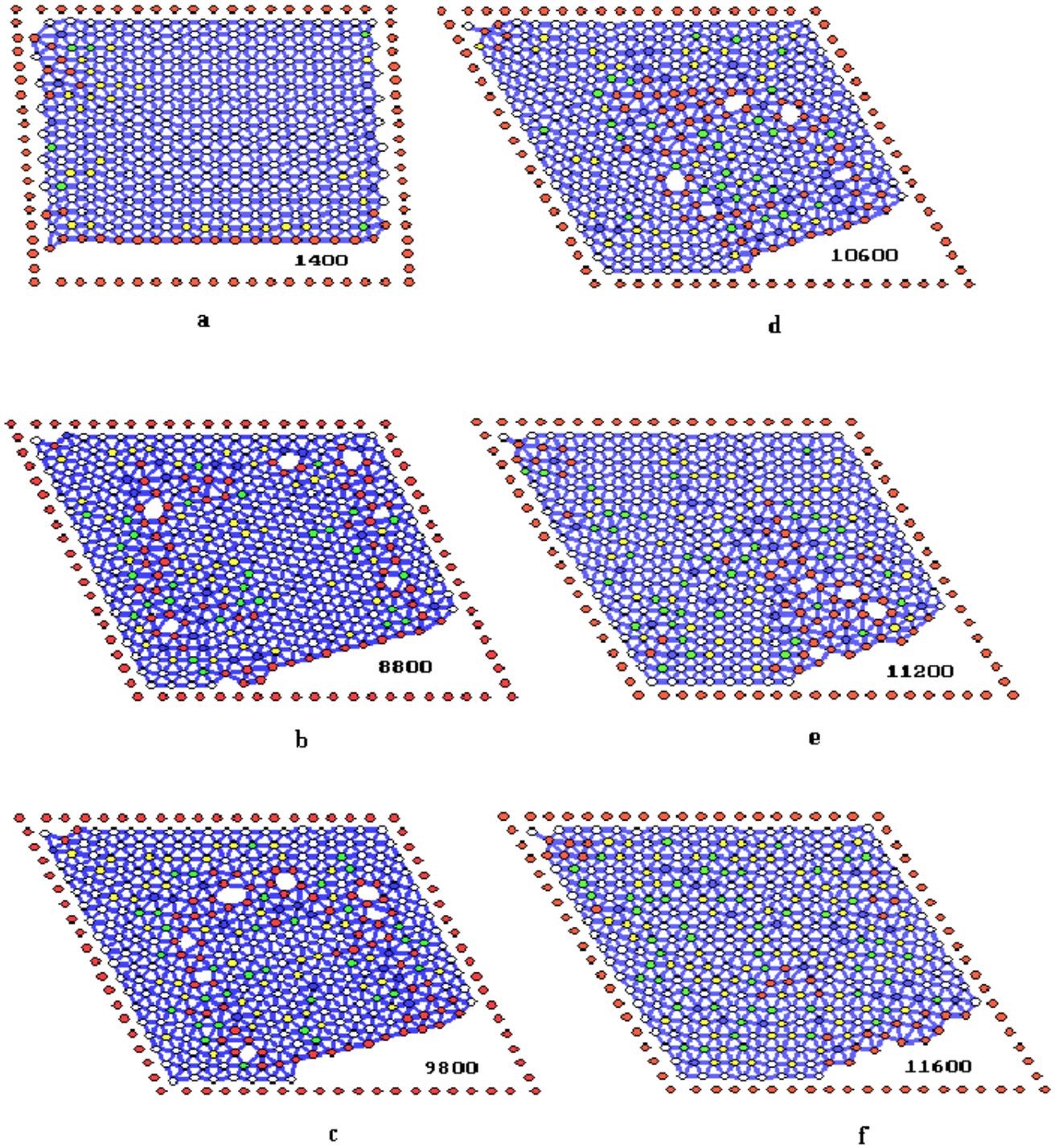}
  \caption{\label{f4}
Phase-like transition with moving boundary consisting from
bubble-bidislocations having pentagon form.} \label{f4}
\end{figure*}

With my point of view,  the defects described above have similar
nature to them considered in ref. \cite{EdOa89} - \cite{Lem02}.
Our "STZs" differ from \cite{EdOa89} - \cite{Lem02} ones because
the former are observed in ideal crystallite, but no in amorphous
metallic glass.

\section{Summary}

    Thus, quark-like behavior of bidislocations is observed
in different numerical experiments. It is established that the
bidislocations may have at least two different structural states.
The first of them has whole stretched chains, the second has
micropores. The bidislocations with micropores, as a rule, have
pentagon form. The states with crack-like form of bidislocations
are possible too (Fig. \ref{f3}c,d). The transient bent form of
microcrack arises at the decay of dislocation (Fig. \ref{f3}e).
The superhigh-speed soliton-like movement of bidislocations with
micropore is noted.

%\newpage
\bibliography{MDforIPD}% Produces the bibliography via BibTeX.

\end{document}